\documentclass{elsart}
\usepackage{natbib}
\usepackage{revsymb}
\usepackage{amssymb}

\def\pole{\alpha_e}
\def\polm{\alpha_m}
\def\Vatat{V_{\mathrm{AtAt}}}
\def\Vatel{V_{\mathrm{AtEl}}}
\def\Vionel{V_{\mathrm{IonEl}}}
\def\lamC{\lambdabar_\mathrm{C}}

\begin{document}

\begin{frontmatter}


\title{Casimir effects in atomic, molecular, and optical physics}


\author{James F. Babb}

\address{ITAMP, Harvard-Smithsonian Center for Astrophysics, MS 14, 60 Garden St., Cambridge, MA 02138}
\tableofcontents

\begin{abstract}
The long-range interaction between two atoms and the long-range
interaction between an ion and an electron are compared
at small and large intersystem separations.
The vacuum dressed atom
formalism is applied and found to  provide a
framework for interpretation of the similarities between
the two cases.
The van der Waals forces or Casimir-Polder potentials are used to obtain insight
into relativistic and higher multipolar terms.
\end{abstract}

\end{frontmatter}

\label{}

\section{Introduction}
\label{sec:intro}
Distance changes everything.  
The same is the case for electromagnetic interaction potential energies between polarizable 
systems.
In atomic, molecular, and optical physics the small 
retarded van der Waals (or Casimir-Polder) potentials between  pairs of 
polarizable systems (either of which is an 
atom, molecule, surface, electron, or ion) for separations
at long-ranges where exchange forces are negligible have been well-studied
theoretically. 
There are also three- and higher-body potentials, cf.~\citep{Sal10},  antimatter
applications~\citep{VorFroZyg05}, and more.
Much of the current interest  in the Casimir interactions
between atoms and walls is due to interests
related to 
nanotechnologies, cf.~\citep{CapMunIan07}, and related attempts to engineer
repulsive forces at  nano-scales~\citep{Mar09}.
Numerous topical surveys and reviews, monographs, book chapters, conference
proceedings, and popular texts 
touching on  particular pairwise potentials are in print---literally a ``mountain of available information''~\citep[p.185]{BonKre97}---and 
many sources contain extensive bibliographies.
It is not uncommon to  run across   statements 
indicating that there has been  a rapid increase in
the number of available papers related to the Casimir effect.
Even a recent book, \textit{Advances in the Casimir effect}~\citep{BorKliMoh09},
focusing mainly on recent results, 
comes  to over 700 pages.

This paper is concerned with bringing to light 
some connections
between theoretical results
from various formulations for 
the zero temperature limits of interactions between 
ground state atoms, ions, or molecules.
The case has been advanced that the practical relevance 
of zero temperature results is questionable, see~\citet{WenDaiNin99},
and although the case is reasonable, you have to start somewhere.
The present article is therefore more selective than comprehensive
and it is organized as follows.
In Sec.~\ref{sec:micromacro} the microscopic and macroscopic natures
of Casimir effects are very briefly surveyed
and the interaction between two atoms is reviewed in 
Sec.~\ref{sec:relativistic} including discussion of the terms of relativistic 
origin  arising for small atomic separations.
In Sec.~\ref{sec:yetanother} the change in the form of
the interaction, when one of the two polarizable systems is charged,
is studied.
The vacuum dressed atom approach is introduced and applied to the case of an electron and an ion in Sec.~\ref{sec:dressed} and it is used in
Sec.~\ref{sec:trivial} to  gain insight into the origin of the numerical factor
``23'' in expressions for potentials related to the Casimir effect.
Finally in Sec.~\ref{sec:multipoles} the treatment of  multipoles
beyond the electric dipole is discussed for two atoms and for an
electron and an ion.
\section{What's a micro effect; what's a macro effect?}
\label{sec:micromacro}

The picture of two well-spaced systems interacting through fluctuating electromagnetic
fields can describe many phenomena. 
The usual definitions are that 
the
\textit{Casimir-Polder} potential~\citep{CasPol48}  is the retarded interaction between
two atoms or an atom and a wall and 
a \textit{Casimir effect}~\citep{Cas48} is the
``observable non-classical electromagnetic force of attraction between two parallel conducting plates''~\citep{Sch75}.
\citet[p.3]{Mil01} traced the change in Casimir's perspective from action
at a distance~\citep{CasPol48} to the local action of fields~\citep{Cas48} or
an equivalence in physical pictures of fluctuating electric dipoles or fluctuating 
electric fields.
The conceptual
realizations of the Casimir effect and of the Casimir-Polder potential have been extended
well beyond their original theoretical models; an
extensive tabulation can be found in~\citep{BuhWel07}.
The term Casimir effect will be
used rather more loosely in the present work recognizing in advance the connection
already established  in the literature with 
the more general pictures of  
``dispersion forces''~\citep{MahNin76} or ``van der Waals forces''~\citep{BarGin84,Par06}.
Also as noted
by  \citet{Bar99} ``By tradition, `Casimir effects' denote \textit{macroscopic}
forces and energy shifts; yet for connected bodies the macroscopic must
be matched to \textit{microscopic} physics, and no purely 
macroscopic model can be guaranteed in advance to reproduce
the results of this matching adequately for whatever purpose is in hand.''
And
as \citet{BarGin84} write,``The fluctuation nature of van der
Waals forces for macroscopic objects is largely the same as for individual atoms
and molecules. The macroscopic and microscopic aspects of the
theory of van der Waals forces are therefore intimately related.''
Moreover, there are  macroscopic formulations that 
can yield results for microscopic systems by taking various limits~\citep{MilLer92,SpaDalMil07,BuhWel07}, but local field corrections
require close study~\citep{HenBoeWil08}.

For the present purposes, the concern is largely with pair-wise potentials and 
their comparison with results from various approaches. 
Atomic units with $\hbar=e=m_e = 1$ are used throughout,
wherein the fine structure constant is $\alpha=1/c$, though
for some formulae $\hbar$ and $c$ are restored. 
It is useful to define the reduced Compton wavelength of the
electron $\lamC \equiv \hbar/m_e c$.
The notational convention of \citet{TikSpr93a} is followed where the
subscripts $\mathrm{At}$, $\mathrm{Ion}$, and $\mathrm{El}$ denote, respectively,
an atom, ion, and electron.

The Casimir-Polder potential for the interaction between two identical atoms is written
as~\citep{CasPol48}
\begin{equation}
\label{CP}
\Vatat(r) = -\frac{1}{\pi r^6} \int_0^\infty d\omega\; \exp (-2\alpha\omega r)
 [\pole (i\omega)]^2 P(\alpha \omega R),
\end{equation}
with 
\begin{equation}
P(x)=x^4+2x^3+5x^2+6x+3,
\end{equation}
and where the dynamic (frequency dependent) electric dipole polarizability is
\begin{equation}
\pole(\omega) = \sum_u \frac{f_{u}}{E_{u0}^2 - \omega^2} ,
\end{equation}
$f_{u}$ is the electric dipole  oscillator strength from the ground
state $0$ to the excited state $u$, $E_{u0}=E_u-E_0$ is the energy
difference, the summation includes an integration over continuum states, and
$\omega$ is the frequency.
An alternative form of Eq.~(\ref{CP})~\citep{Boy69,SprKel78} is
\begin{equation}
\label{CP-trig}
\Vatat(r) = -\frac{\hbar c}{\pi} 
\lim_{\mu\rightarrow 0}
\int_0^\infty dk\;  k^6 e^{-\mu k}
 [\pole (\omega)]^2 I(kr),
\end{equation}
with $\omega=kc$ and 
where
\begin{equation}
I(x) = \sin(2x)(x^{-2}-5x^{-4}+3x^{-6}) + \cos(2x)(2x^{-3}-6x^{-5}) .
\end{equation}

The interaction potentials  given in Eqs.~(\ref{CP})
and (\ref{CP-trig}) are valid for all separations larger than 
some tens of $a_0$,
and   do not take into account electron charge cloud overlap, spin, magnetic susceptibilities, 
for example, 
though these have all
been studied.
The interaction potential $\Vatat(r)$, given by either 
Eq.~(\ref{CP}) or (\ref{CP-trig}), does contain the van der Waals interaction, certain relativistic effects, and higher
order effects, as well as the asymptotic form first obtained by~\citet{CasPol48},
\begin{equation}
\label{CP-asymp}
\Vatat(r) \rightarrow -23 \frac{\hbar c}{4\pi r^7} [\pole(0)]^2 , \qquad r\rightarrow\infty.
\end{equation}
For the  hydrogen atom, $\pole(0)=\frac{9}{2}$.

\section{Relativistic terms}
\label{sec:relativistic}

Before studying the long-range Casimir-Polder interaction potential in detail, it is useful to look
at  the ``small $r$'' expansion\footnote{The term ``short-range'' is avoided and reserved for exchange,
overlap, and forces that
are, for example,
exponentially decaying~\citep{BarGin84}.
Thus, the term ``long-range'' interaction potential here will indicate the
form valid for 
intersystem separations, typically from several to tens of $a_0$
to infinity,  
such as those in Eq.~(\ref{CP}) and (\ref{CP-trig}), which
have a ``small $r$'' expansion, Eq.~(\ref{dip-dip}) and a ``large $r$''
expansion~Eq.(\ref{CP-asymp})} of the full potential Eq.~(\ref{CP})
for  distances, say, of the order $20\;a_0$.

Expanding Eq.~(\ref{CP}) for small $r$,
the potential is 
\begin{equation}
\label{dip-dip}
\Vatat(r) \sim -\frac{C_6}{r^6} +  \alpha^2 \frac{W_4}{r^4} 
   + \mathcal{O}(\alpha^3/r^3), \qquad r\sim 20\;a_0 .
\end{equation}
The first term in the expansion is
the van der Waals potential with van der Waals constant,
\begin{equation}
C_6 = \frac{3}{\pi} \int_0^\infty  d\omega\;  [\pole(i\omega)]^2 ,
\end{equation}
and 
for two H atoms,  $C_6 = 6.499\,026\,705\,405\,84$~\citep{Wat91}.

The term of order $\alpha^2$ relative to the 
van der Waals potential  is a relativistic  correction
\begin{equation}
\label{W4}
W_4 = \frac{1}{\pi}  \int_0^\infty d\omega\;  \omega^2 [\pole(i\omega)]^2, 
\end{equation}
which can be traced
back~\citep{PowZie57a} to the ``orbit-orbit'' effective potential appearing in the Breit-Pauli
reduction of the Dirac equation~\citep{MeaHir66}.

The numerical value of $W_4$ for two H atoms is
$0.462\,806\,538\,843\,273$ according to \citet{Wat91},
who used a momentum space approach and expansion in 
Pollaczek polynomials; he also obtained the  highly accurate
value of $C_6$ quoted above.
Certain exact representations of the dynamic polarizability function of H 
also facilitate evaluations of $W_4$~\citep{DeaYou71} and of $C_6$~\citep{OCaSuc68}.

Small  relativistic terms were applied  in a few cases to 
potential energy functions of light diatomic molecules,
see for example \citet{PrzCenKom10}, and  
where improved accuracies were sought for precision calculations, for example, 
such as those of low-energy ultra-cold atomic collisions~\citep{ZygDalJam03}
or of the ionization potential of the hydrogen molecule~\citep{PisLacPrz09}.

Recently~\citet{Pac05} reanalyzed the  Casimir-Polder potential 
complete to terms of $\mathcal{O}(\alpha^2)$, but  expressed 
it in such  a way that
its form is valid over all distances sufficiently large that
the atomic wave functions don't overlap; not just in the large $r$  limit.

\section{Yet another repulsive interaction}
\label{sec:yetanother}

In the previous section,
the original Casimir-Polder potential was introduced and seen to be
attractive, but there are several known cases where repulsive potentials
have been predicted theoretically.\footnote{V.~Hushwater,
\textit{Survey of Repulsive Casimir Forces,} unpublished talk, 
ITAMP Casimir workshop, Cambridge, MA, 
November 16, 2002.}
Thus,  \citet{FeiSuc68,FeiSuc70} used a general dispersion-theoretic scattering approach  to 
show that the potential given in~Eq.~(\ref{CP-asymp}) can be generalized for two systems $A$ and $B$ to an expression bilinear
in the electric and the magnetic polarizabilities of each system,
\begin{eqnarray}
\label{FS-full}
\Vatat(r)  \sim & & -\frac{\hbar c}{4\pi r^7} [23(\pole^A\pole^B + \polm^A\polm^B)  \nonumber \\
  &   &  
     -7 (\pole^A \polm^B + \pole^B\polm^A)] + \mathcal{O}(r^{-9}) , \qquad r\sim\infty
\end{eqnarray}
where $\pole^A$ and $\pole^B$
are the static polarizabilities $\pole(0)$, respectively, of $A$ and $B$, 
and  $\polm^A$ and $\polm^B$, with $\polm\equiv\polm(0)$,
are, respectively, the static magnetic polarizabilities of $A$ and $B$.
Note that  the cross term in the potential containing 
the product of  $\pole$ and $\polm$ leads to a repulsive force.\footnote{The $\pole^A \polm^B$ term supports the result that 
the interaction between two plates is  repulsive,
if one plate  $(A)$ has only 
infinite permittivity and one plate $(B)$ is only  infinitely permeable~\citep{Boy74};
an alternative argument not making use of Eq.~(\ref{FS-full}) is given 
by \citep{SchSpr98}.}
More detailed discussions concerning the treatment of the magnetic terms for the interaction
between two atoms can be found in~\citet{Sal00c,Sal10}.

There is  another repulsive Casimir-Polder potential, perhaps not as well known.
For the  scattering interaction
between a charged, structureless particle $B$ and  a neutral
polarizable particle $A$~\citep{BerTar76,SprKel78} or for the interaction 
between a charged, structureless particle $B$ 
and an ion $A$~\citep{KelSpr78,SprKel78},
there is an interaction (for $B$ an electron) given by
\begin{equation}
\label{KSBT-EM}
\Vatel (r) \sim \Vionel (r) \sim \frac{\lamC e^2}{4\pi r^5}
  [  11\pole^A+ 5 \polm^A ], \qquad r\sim\infty .
\end{equation}
For either of the the two cases (the target is
neutral or it is charged) 
the asymptotic result Eq.~(\ref{KSBT-EM}) applies, 
but the complete potentials including other corrections 
are not identical,  due to the remnant
$1/r$ Coulomb interaction in the charged particle and ion case higher order corrections
at large $r$  differ, as  emphasized by~\citep{Au86,Au89}.

The long-range Casimir potential $\Vionel (r)$ is the present
object of interest but it is useful sometimes to write the full
potential with ``instantaneous'' Coulomb interactions as well.
Therefore, the full potential $U(r)$, including the charged particle
and ion electric interactions (but neglecting the 
dominant $1/r$ Coulomb
potential), is at large $r$ 
\begin{equation}
\label{ION-LONG}
U(r) = 
-\frac{1}{2} e^2 \pole r^{-4}
+ \frac{11}{4\pi} \lamC e^2 \pole r^{-5} ... \qquad r\sim\infty ,
\end{equation}
where $\pole$ is the polarizability of the ion.
The first term in Eq.~(\ref{ION-LONG})  is the polarization potential.
The second term is the asymptotic Casimir-Polder-type interaction, which 
was confirmed theoretically~\citep{FeiSuc83,Au86}.
The  Casimir-Polder 
potential for the interaction
between an electron and an ion was expressed in a fashion similar to the result
for two atoms given in Eq.~(\ref{CP})
by \citet{AuFeiSuc84} using a dispersion-relation analysis,
and later using Coulomb gauge,  old-fashioned perturbation theory by 
\citet{BabSpr87} and \citet{Au88}.

It can be expressed~\citep{BabSpr87} in the compact form similar to Eq.~(\ref{CP-trig}),
\begin{equation}
\label{ION-ALL}
\Vionel (r) = \frac{e^2}{\pi}\lamC  \lim_{\mu\rightarrow 0}
\int_0^\infty dk\;  e^{-\mu k}
k^4 F(k) I(kr),
\end{equation}
where
\begin{equation}
F(k) =  \sum_u f_u /[E_{u0}(E_{u0}+E_k)] .
\end{equation}

Taking account of the Coulomb interactions,
the potential has the expansion for small $r$,
\begin{equation}
\label{KSBT-short}
U(r) + \Vionel (r)
 \sim  -\frac{1}{2} e^2 \pole r^{-4} +  3 \beta  r^{-6} + ( \alpha^2/Z^2) r^{-4}
 ...,\quad r\sim \textrm{a few} \;a_0,
\end{equation}
where $Z$ is the charge of the ion,
and $\beta=\textstyle{\frac{43}{8}}Z^{-6}$~\citep{KleHahSpr68,DalDraVic68}.
Note the disappearance for small $r$ of the $r^{-5}$ term that was present 
in the large $r$ potential, Eq.~(\ref{ION-LONG}).

It was emphasized by \citet{KelSpr78b} and \citet{FeiSucAu89}
that  that the $3\beta r^{-6}$ term disappears at large $r$. 
Thus, the potential can be written in the form~\citep{BabSpr87}
\begin{eqnarray}
\label{KSBT-full}
\Vionel(r) + 3 \beta  r^{-6} & =&
-\frac{\hbar c}{\pi}
\lim_{\mu\rightarrow 0} \int_0^\infty dk\; k^6 e^{-\mu k} I(kr)  \nonumber \\
& &\times  \{\pole(k)\alpha_\mathrm{out}(k) - 2\frac{\hbar c}{e^2} \alpha_\mathrm{out}(k)[\beta(k)-\beta]\},
\end{eqnarray}
using 
\begin{equation}
\label{alpha-out}
\alpha_\mathrm{out}(\omega) = -e^2/m\omega^2 ,
\end{equation}
\begin{equation}
F(k) = \pole(k) - 2 \frac{k}{\alpha}\beta(k) ,
\end{equation}
and 
\begin{equation}
\beta(k) = \frac{1}{2} \sum_u f_u /[E_{u0}(E_{u0}^2 + E_k^2)].
\end{equation}

The 
effective potential Eq.~(\ref{ION-ALL}) was evaluated
numerically and used to study theoretically the 
energy shift arising from the 
interaction between an  electron bound in a high Rydberg state $|ln\rangle$  with 
principal quantum number $n$,
and with angular momentum $l \sim n$.
Experiments on highly excited $n=10$ Rydberg states
of He were carried out ands several theoretical formulations were developed.
The  details about experiments and theories, and their comparisons,
are very completely presented in the book edited by~\cite{LevMic92},
see also~\citep{Hes92,SteLun00,Dra93,Lun05}. Many other terms must be considered carefully
in the theoretical calculations; the details don't directly relate further though to the present
paper.

Recently two groups have reproduced the electric ($\pole$) 
part of Eq.~(\ref{KSBT-EM})
using  different arguments, but both  approaching  the interaction
between a 
charged particle  and a neutral particle as one loop quantum field
theoretic calculations.
\citet{PanWidSri90} obtained Eq.~(\ref{KSBT-EM}) and 
and  traced the  $r^{-5}$ result result back to a change in the mass 
induced by 
``condensed-matter  renormalization'' of the  electromagnetic fluctuations~\citep{PanWid94}.
The repulsive potential is attributed to soft-photon infrared renormalization.
In contrast, \citet{HolDon04}  were seeking to identify
cases where classical effects are found within one-loop diagrams.
Using an effective field theory approach, they find quantum corrections
to the classical polarization potential; their result is identical
to the $\pole$ part of Eq.~(\ref{KSBT-EM}). They identify the $r^{-5}$ term as
a quantum correction to the polarization potential,
which arises from the infrared behavior
of the Feynman diagrams, when at least two massless
propagators occur in a loop contribution.

In a subsequent study,
\citet{Hol08}  obtained~Eq.~(\ref{KSBT-EM}) and  observed that the 
$r^{-5}$  term in Eq.~(\ref{KSBT-EM}) might be 
associated with zitterbewegung and he 
noted  that, under the influence of this effect, 
``in the quantum mechanical case
the distance between two objects is uncertain by an amount 
of  order  the Compton wavelength due to zero
point motion,''
$\delta r \sim \lamC$, hence
\begin{equation}
V(r) \sim \frac{1}{r^4} \rightarrow  \frac{1}{(r\pm \delta r)^4} \approx 
\frac{1}{r^4} \mp 4\lamC\frac{1}{r^5} .
\end{equation}
It is an intriguing argument, though the ambiguity of the sign is unresolved.
Zitterbewegung would normally be attributed to virtual electron-positron
transitions~\citep[p.~322]{Mil94} at the length scale less than $\lamC$, which would seemingly
place the effect outside of the realm of the low energy fluctuation arguments
used by, for example,~\citet{SprKel78} in deriving Eq.~(\ref{KSBT-EM}).
Nevertheless, the calculations of \citet{HolDon04} are concerned
with large $r$, so there must be a connection to the scale of $\lamC$
and this will be addressed in the next section.

The  effective field theory of \citet{Hol08}
allows the longest-range parts of electromagnetic scattering processes to be isolated,
and he extended the asymptotic (large $r$) 
results for the interactions between
two systems, with and without spin, to the case
where  one or both systems are electrically neutral;
see also~\citet{SucFei-LRF}.

\section{Nonrelativistic molecules and dressed atoms}
\label{sec:dressed}

In the theoretical   ``vacuum dressed atom" approach, a ground state
``bare'' atom interacts with the vacuum  electromagnetic field.
The combination system of the atom and the field  is taken to
be in the lowest possible energy state 
of the noninteracting atom-field system, and the
zero-point fluctuations of the field are seen
as  inducing virtual absorption and re-emission of photons
in the atom---the ``vacuum dressed atom'' is then 
the  system comprised
of the atom and the 
associated cloud of virtual photons~\citep{ComPalPas95,ComPasPer95}.
A good account of the development of the concept of atoms dressed by 
the vacuum electromagnetic field
is given by~\citet{ComPasPer06}.
The energy density can be calculated and  used to obtain expressions
for long-range potentials and other physical quantities,
as was shown 
in quantum optics~\citep{ComPasPer95,CirPas96,Sal10}
for  the nonrelativistic free electron interacting with the vacuum
electromagnetic field and the nonrelativistic hydrogen atom.

\citet{ComSal91} [see also \citet{ComPasPer95}] considered a slow electron 
interacting with the vacuum field.
The key observation is that the cloud around the electron is due to emission
and reabsorption of virtual photons in the course of recoil events.
They point out the zitterbewegung due to relativistic fluctuations would enter and contribute
a cloud of size of order $\lamC$. This effectively
limits their nonrelativistic model to distances $\gg\lamC$, so that
the positron cloud can be neglected and the 
electron has the physical charge $-e$, accordingly, only low frequency photons enter.
In this picture the virtual cloud affects the
field surrounding the charge and changes the average values of the  squares of the
electric and magnetic fields.

They calculate the classical and quantum contributions to the
energy density around the electron  both moving and at rest and 
for the electron at rest, they find (for $r\gg\lamC$) 
\begin{equation}
\label{electron-e}
\langle E_\mathrm{e} (r) \rangle = \frac{e^2}{8\pi r^4}
  +\frac{5}{16\pi^2} \frac{e^2}{r^4}\frac{\lamC}{r}
+ \frac{\hbar c}{4 V} \sum_k k
\end{equation}
and
\begin{equation}
\label{electron-m}
\langle E_\mathrm{m} (r) \rangle =-\frac{5}{16\pi^2} \frac{e^2}{r^4}\frac{\lamC}{r}
+ \frac{\hbar c}{4 V} \sum_k k  ,
\end{equation}
where $E_\mathrm{e} (r)$ and $E_\mathrm{m}(r)$ are, respectively,
the electric and magnetic energy densities at a distance $r$ from
the electron, $V$ is the quantization volume,
and the sum is over the vacuum field modes.

For the present purpose, one of their key observations is the appearance
of the $r^{-5}$ contribution which \citet{ComSal91} attribute to the virtual photon
cloud  surrounding the electron fluctuations
that arise due to interference between the virtual photons emitted and
absorbed by the electron and zero-point  field fluctuations.
The $r^{-5}$ term is deemed purely quantum in nature~\citep{ComPasPer95}.
The energy densities can be directly related to the Casimir-Polder potential,
as noted by \citet{PasPow87}.
What is striking is the  similarity in form between
Eqs.~(\ref{electron-e}) and (\ref{electron-m}) 
and the large $r$  potential for the interaction of 
an electron and an ion, Eq.~(\ref{ION-LONG}).
Evidently, both the classical polarization potential and the retarded
asymptotic Casimir-Polder potential are present.

As discussed above, 
\citet{HolDon04}  showed that, within a diagrammatic,
effective field theory approach, classical effects
can arise.  In particular, the energy density of a particle in 
a plane wave calculated 
by~\citet{HolDon04} agrees in form with the dressed electron result containing
both a $r^{-4}$ polarization potential and the ``purely quantum'' $r^{-5}$ asymptotic retarded potential.

For the nonrelativistic hydrogen atom, the analysis was carried out
again within the vacuum dressed atom formalism;
the extensive calculations can be found in~\citep{PasPow87,ComPasPer95,ComPasPer06}.
The analysis is complicated, but it is similar to that carried out by \citet{BabSpr87}
and \cite{Au89}.
For example, Eq.~(7.148) of \citet{ComPasPer95}  describes
the longitudinal electric field and transverse electric field
contributions to the energy density around a hydrogen atom,
\begin{equation}
 \frac{1}{4\pi}  {}^ \prime\langle\phi | E_{||}(\mathbf{x}) - E_{\perp}(\mathbf{x}) |\phi\rangle {}^\prime
 \sim \frac{1}{r^3} \int \frac{k^2 j_2(kr)}{\omega_N + \omega_k} dk 
\end{equation}
and it 
is almost identical to $V_{IT}$, Eq.~(4.16) found by \citet{BabSpr87} for the
contribution of one instantaneous Coulomb photon and one transverse
photon to the effective potential in the 
case of an electron and an ion.

In an earlier study using the 
virtual photon cloud picture, \citet{PasPow87}
note  that the $r^{-6}$ term in the description
of the energy density around a ground state
hydrogen atom disappears at large $r$  similarly to 
the way the van der Waals form $r^{-6}$
form is replaced at asymptotic distances
by the $r^{-7}$ form.
They  find that nonretarded effects of order $\sim  r^{-6}$
in the energy density are absent in the far zone of the hydrogen atom
and they obtain the simple form for the energy density
with 
an $\mathcal{O}(r^{-7})$ term related to the virtual charge cloud,
\begin{equation}
  \frac{1}{8\pi} 
\langle\Psi | {
F^2  } | \Psi\rangle  -E_{zp} = 
\frac{23}{16\pi^2} \frac{\hbar c}{r^7}\pole ,
\end{equation}
where $F$ is the electric field.

\citet{Rad90a} carried out a relativistic calculation of the electromagnetic
virtual cloud of the ground state hydrogen atom using a Dirac formalism.
According to~\citet{ComPasPer06} his work was supposed to be an independent
calculation of the energy density of the vacuum dressed hydrogen atom.
For the energy density
due to the electric field,  his  result  in the large $r$ limit is  
\begin{eqnarray}
\label{dirac}
\frac{1}{2}
\langle F^2(\mathbf{r}) \rangle =  
\frac{13\hbar ce^2}{16\pi^2} &&  \sum_n \frac{1}{E_{n1}} \langle 1|x|n\rangle\langle n |x|1\rangle \frac{1}{r^7} \nonumber \\
&&+ \frac{7\hbar ce^2}{16\pi^2} \sum_n \frac{1}{E_{n1}} \langle 1|x_i|n\rangle\langle n |x_j|1\rangle \frac{\hat{r}_i\hat{r}_j}{r^7} .
\end{eqnarray}
Identifying  the tensor electric dipole polarizability in Eq.~(\ref{dirac})
\begin{equation}
2\sum_n \frac{1}{E_{n1}} \langle 1|x_i|n\rangle\langle n |x_j|1\rangle = \alpha_{e,ij},
\end{equation}
Eq.~(\ref{dirac}) agrees with the two-level ``Craig-Power'' model~\citep{ComPasPer95}
result for the energy density in the large $r$ limit,
\begin{equation}
\label{Craig-Power}
\langle E_e(\mathbf{r}) \rangle = \frac{1}{32\pi^2} \hbar c
\alpha_{ij} (13\delta_{ij} + 7 \hat{r}_i\hat{r}_j)\frac{1}{r^7}.
\end{equation}

In an unrelated study of the
Casimir-Polder potential for an electron interacting with a hydrogen
molecular ion core,
\cite{BabSpr94}  obtained an expression almost identical to Eqs.~(\ref{dirac})
and (\ref{Craig-Power}). The tensor polarizability arises from 
the anisotropic interaction arising from the cylindrical symmetry 
of the diatomic molecule core.

\citet{ComPasPer95} interpret the large $r$ Casimir-Polder potential
as the interaction between the vacuum dressed  ``source'' atom with polarizability 
$\alpha_S$ and the ``test'' atom with polarizability $\alpha_T$ 
\begin{equation}
V(r) = \frac{-23}{4\pi} \hbar c \alpha_S \alpha_T \frac{1}{r^7}
    = -4\pi \alpha_T\langle E_e^S (\textbf{r})\rangle ,
\end{equation}
where the energy density is generated by the source at point $\mathbf{r}$
in the absence of the test atom.

Another interesting point emphasized by~\citet{ComPasPer95} is that
\begin{equation}
V(\mathbf{r}) = -4\pi\alpha_T \langle E_e^S (\textbf{r})\rangle ,
\end{equation}
``thus the van der Waals force provides a means of measuring directly the
electric energy density of a source both in the near and in the far regions.''

\section{Not a trivial number}
\label{sec:trivial}

In his contribution to the proceedings of a  conference held in 
Maratea, Italy,\footnote{June 1--14, 1986}
\citet{Cas86} wrote,``In the theory of the so-called Casimir effect two lines of approach 
are coming together. The first one is concerned with Van der Waals forces, the
second one with zero-point energy."
Today, that connection is well-established, though the
``reality'' of zero-point energy is still debatable; see
the very accessible article by~\citet{RugZinCao99}, and also~\citet{Jaf05}.

In Eq.~(\ref{CP-asymp}), it was shown that the asymptotic potential
for the interaction between two electrically polarizable particles contains 
the factor 23, as does the asymptotic potential for two magnetically 
polarizable particles, see Eq.~(\ref{FS-full}).
The factor 23 has reappeared in other situations. In the 
asymptotic interaction between
and electron and an ion,
expanding Eq.~(\ref{KSBT-full}) for large $r$ and keeping one more
term past that given in Eq.~(\ref{ION-LONG}),~\citep{FeiSuc83},
the potential is 
\begin{eqnarray}
\label{ion-asymp}
U(r) & \sim & -\frac{1}{2} e^2 \pole r^{-4}+ \Vionel(r)  \nonumber \\ 
 && \sim-\frac{1}{2} e^2 \pole r^{-4} 
+ \frac{11}{4\pi}\lamC e^2\frac{\pole}{r^5} + \frac{23}{4\pi}
\frac{e^2 \lamC}{a_0^2}\frac{\gamma(0)}{r^7} + ... ,\quad r\sim\infty .
\end{eqnarray}

According to \citet{FeiSucAu89}, when told of this result [that is,
Eq.~(\ref{ion-asymp})], at the Maratea conference,
Casimir replied,
``23 is not a trivial number. I am happy to see that.''
However, the appearance of the 23 in a way nearly identical to the
result for the asymptotic atom-atom 
potential was not explained completely~\citep{FeiSuc83}.
In addition, the sign of the term containing 23 is opposite to that for the case of
two atoms.

Evidently, while the complete potential for the electron-ion case can be
expressed as Eq.~(\ref{ION-ALL}), expansion for large $r$  yields the two
terms of Eq.~(\ref{KSBT-EM}).
I conjecture that 
the $\mathcal{O}(r^{-5})$ term can be interpreted as
the effective potential arising from the energy density of
the weakly bound, vacuum dressed, electron ``source''
interacting with the ion core ``test'' $\pole$,
\begin{equation}
\pole \langle E_\mathrm{out}^2  \rangle \sim \pole ( \frac{1}{r^4} + \lamC \frac{1}{r^5} ),
\end{equation}
in accord with the ideas of~\citet{ComSal91},
supported by the large $r$ one loop calculations of~\citet{HolDon04}
and \citet{Hol08}.
The other term $\sim r^{-7}$  can
be interpreted as the source term of the fluctuating 
vacuum dressed polarizable ion core ``source'' acting on the electron
``test'' particle,
\begin{equation}
\label{free-source}
-4\pi\alpha_\mathrm{out}(\bar{\omega})\langle \bar{E}_e^S (\textbf{r})\rangle \sim 
-\alpha_\mathrm{out}(\bar{\omega} )\frac{23}{4\pi} \hbar c \gamma r^{-7},
\end{equation}
where $\bar{E}_e^S$ indicates that the electric field energy density
is modified due to the Coulomb binding
and $\bar{\omega}$ is a characteristic energy.
This is to be expected based on arguments given by~\citep{Au86,Au89}
for the Rydberg helium case and by \citet{ComSal91} for the vacuum dressed slow electron
and vacuum dressed hydrogen atom.
Using Eq.~(\ref{alpha-out}) for $\alpha_\mathrm{out}(k)= - \lamC \alpha k^{-2}$,
where $kc=\omega$,
evaluated at $\bar{\omega}=  \alpha c /a_0$~\citep{FeiSuc83}
Eq.~(\ref{free-source})  yields a term
in general agreement with Eq.~(\ref{ion-asymp}),
\begin{equation}
-4\pi\alpha_\mathrm{out}(\bar{\omega})\langle \bar{E}_e^S (\textbf{r})\rangle \sim 
   +    \lamC \frac{23}{4\pi} \gamma r^{-7}    .
\end{equation}

The approach of adding the two interactions is consistent with the interpretation
of the fluctuating field approach to Casimir-Polder interactions proposed 
by~\citet{PowThi93b}. Namely, that the dipole in each particle
is induced by the vacuum fluctuations of the electromagnetic field.

\section{Reconciling multipoles}
\label{sec:multipoles}
\subsection{Two atoms}

The extension  beyond 
electric and magnetic dipoles
for the retarded van der Waals (or Casimir-Polder) potential
between two neutral spinless systems was achieved by~\citet{AuFei72}.
Using  scattering analysis they were able to obtain 
integral forms for the complete potential for each multipole, valid
for all separations greater than some tens of $a_0$, and they gave the  
first several terms in each of the 
small $r$ and large $r$ expansions
of the potentials.
Their result, as noted by~\citet{Fei74}, and
as emphasized by~\citet{PowThi96}, included the property ``that the
expansions of the electric (magnetic) form factors include high-order magnetic
(electric) susceptibilities in addition to electric (magnetic) polarizabilities.''

The first several electric multipole results of \citet{AuFei72} were used 
for applications to calculations of  the binding energy of  the helium dimer by
\citet{LuoKimGie93} and by  \citet{CheChu96}  
and for applications to 
ultra-cold atom scattering by~\citet{MarBabDal94}, who evaluated 
the expressions for a pair of hydrogen atoms and for a pair of like alkali-metal atoms.

A few years later, in a thorough analysis, \citet{SalThi96} and \citet{PowThi96} pointed 
out that the results of the~\citet{AuFei72}
analysis did not concur with other results  obtained  
evidently independently by~\citet{JenSalThi94b} who
used a different approach.
\citet{PowThi96} argued that the correct form for the next
order Casimir-Polder interaction, arising from the interaction
of an electric dipole and an electric quadrupole is 
\begin{equation}
\label{dip-quad}
V_{12} (r) =  - \frac{1}{3\pi} \int_0^\infty d\omega \exp(-2\alpha\omega r)
 \pole(i\omega) \alpha_{e;2} (i\omega) P_{12}(\omega\alpha r),
\end{equation}
where $\alpha_{e;2}$ is the electric quadrupole polarizability and 
\begin{equation}
P_{12}(x) =\textstyle{ \frac{1}{2}}x^6 + 3 x^5 + \textstyle{\frac{27}{2}}x^4 + 42x^3
 + 81 x^2 +90x + 45 .
\end{equation}

For small $r$,
\begin{equation}
V_{12} (r) \sim -\frac{C_8}{r^8}  + \alpha^2 \frac{W_6}{r^6} + ...,
\end{equation}
where
\begin{equation}
C_8 = \frac{15}{\pi}\int_0^\infty d\omega \;\omega^2 \pole(i\omega)\alpha_{e;2}(i\omega), 
\end{equation}
and the coefficient of the relativistic correction is 
\begin{equation}
\label{W6}
W_6  = \frac{3}{\pi}  \int_0^\infty d\omega \;\omega^2 \pole(i\omega)\alpha_{e;2}(i\omega)
\end{equation}

\citet{MeaHir66} obtained the relativistic corrections of relative order $\alpha^2$
for two hydrogen atoms.
For the orbit-orbit interaction $H_{LL;0}$, the corresponding effective 
potential contributions arise as
powers of $r^{-4}$ and $r^{-6}$. Their approach is not valid in the large $r$ limit,
but it should agree with the small $r$ form of $\Vatat(r)$ and $V_{12}(r)$.

In turn~\citet{MarBabDal94} noticed a discrepancy between the result
of Meath and Hirschfelder and \citet{JohEpsMea67} for the $r^{-6}$ relativistic
term for
small distances and the result obtained by expanding the potential of Au and Feinberg
for small $r$.
In contrast, expansion of the  revised dipole-quadrupole 
potential $V_{12}(r)$ of~\cite{JenSalThi94b}
for small $r$ provides a value for $W_6$, see Eq.~(\ref{W6}),
in agreement with the expression of~\citet{MeaHir66} and \citet{JohEpsMea67} 
for the $r^{-6}$ term in the expansion of the Breit-Pauli equation.
Accordingly, earlier results, such as the results
of~\citep{CheChu96} for $W_6$ in 
should  be multiplied by $\frac{3}{2}$.

Later, \citet{MarYou99} rederived the atom-atom potential  accounting
for  magnetic and other terms to higher order, see also~\citep{Sal00c}.
Marinescu and You note that numerically, at least, for their evaluations of
the like alkali-metal atom pairs, that the relative error between
results from the two approaches is smaller than $10^{-5}$.
In any case, other terms, such as mass polarization, Darwin interaction terms, 
and Lamb
shifts,  would have to be included at the correct order for a complete
description.

Asymptotically, for  large $r$ 
\citep{Thi88,JenSalThi94b,YanDalBab97,MarYou99} 
\begin{equation}
\label{V-dq-asymp}
V_{12}(r) \sim -\frac{531 \hbar c}{16 \pi r^9}
\pole(0) \alpha_{e;2}(0) .
\end{equation}

\cite{Fei74} expected that for atoms---with the exception of possibly the
magnetic-magnetic case for two hydrogen atoms, see \citep{FeiSuc68}---magnetic and higher order multipole effects would be negligible
for domains where retardation was important.
Thus, Feinberg was motivated to use the scattering approach to study the case of
two superconducting spheres and he obtained a series in powers of the sphere-sphere
separation distance.

Using a new formalism based on a scattering approach, \citet{Emi08,EmiJaf08}
investigated the Casimir energy between two spheres.
For large separations, they obtain an expansion in the separation
distance $r$.
The lead term is of order $r^{-7}$ and is given by
Eq.~(\ref{FS-full}), where the polarizabilities correspond to
those of the spheres.
Moreover, the  next term, of order $r^{-9}$ is given
by Eq.~(\ref{V-dq-asymp}).
In another calculation using the scattering
approach,~\citet{Emi10} obtains the large
$r$ interaction potential for two anisotropic objects;
his result is in agreement with the earlier results for  
two anisotropic  particles given by~\cite{CraPow69}.

\subsection{An electron and an ion}

Expanding Eq.~(\ref{ION-ALL}) for small $r$,
it was shown above in Eq.~(\ref{KSBT-short}) that there
is a term $( \alpha^2/Z^2) r^{-4}$.
This  relativistic term is identical to the atom-atom case, which
is known to result from perturbation
treatment of the  Breit interaction with the Coulomb interaction~\citep{PowZie57a,Au89}.

Some time after Eq.~(\ref{CP-trig})
was obtained, the complete long-range potential including multipoles for
an ion and a neutral spinless system was obtained by~\citet{FeiSuc83}
and by \citet{Au85}.
According to \citet{Hes92}, the  result of \citet{Au85} for the next term is
\begin{equation}
V_{E,1}(r) = \textstyle{\frac{9}{16}}\alpha^2 r^{-6}.
\end{equation}
\citet{Hes92} carried out a perturbation theoretic calculation of the
relativistic corrections for the ion-electron system, analogously to the calculation of \citet{MeaHir66} for the
atom-atom interaction.
His analysis is in disagreement with \citet{Au85}, but in agreement with comprehensive
calculations by \citet{Drake-LRF}, indicating an unresolved 
discrepancy between the dispersion theoretic result and perturbation theoretic results
at $\mathcal{O}(\alpha^2 r^{-6})$ for small $r$ limit of the the ion-electron system.

\section{Conclusion}
\label{conclusion}

The Casimir effects for the interaction between two atoms and for
the interaction between an ion and an electron were investigated and,
respectively, their expansions lead to asymptotic
terms of order $r^{-7}$ and $r^{-5}$.
The second correction at large $r$ for the ion and electron case is similar
to the leading term at large $r$ for the case of two atoms.
It was shown that the vacuum dressed atom picture provides a framework
for interpretation of this similarity.

Reconciliation of interaction potentials
for electric dipole and electric quadrupole multipoles between atom-atom and ion-electron cases
led to insight concerning a discrepancy  between a scattering dispersion theoretic
calculation and a perturbation theoretic calculation of the ion-electron interaction
for the
electric quadrupole relativistic term.

As interest in the potential applications of Casimir effects
in atomic, molecular, and optical physics increases, limiting results
for interaction potentials at zero temperature---such as
those presented here---may provide useful insights
and checks on calculations for more complicated geometries.
Hopefully, it will be a long time until it is true that \textit{nothing can be added to vacuum studies.}
\section{Acknowledgments}
I am indebted to several  colleagues who have shared their
knowledge with me over
the years on topics related to this article. In particular, Larry Spruch
Alex Dalgarno, Joe Sucher, Akbar Salam, and Peter Milonni provided helpful insights.
ITAMP is partially supported by a grant from the NSF to Harvard University
and the Smithsonian Astrophysical Observatory.



\begin{thebibliography}{87}
\expandafter\ifx\csname natexlab\endcsname\relax\def\natexlab#1{#1}\fi
\expandafter\ifx\csname url\endcsname\relax
  \def\url#1{\texttt{#1}}\fi
\expandafter\ifx\csname urlprefix\endcsname\relax\def\urlprefix{URL }\fi

\bibitem[{Au(1985)}]{Au85}
Au, C.-K., 1985. Diamagnetic form factors in photon-atom scattering and
  higher-multipole {Casimir} effect in heliumlike {Rydberg} ions. Phys. Rev. A
  31, 1310--1318.

\bibitem[{Au(1986)}]{Au86}
Au, C.-K., 1986. Coulomb-gauge electrodynamics analysis of two-photon exchange
  in electron-atom scattering. Phys. Rev. A 34, 3568--3579.

\bibitem[{{Au}(1988)}]{Au88}
{Au}, C.~K., 1988. {Coulomb-gauge electrodynamics analysis of two-photon
  exchange in electron-atom scattering. II. Interaction at all distances beyond
  atomic dimensions}. Phys. Rev. A 38, 7--12.

\bibitem[{Au(1989)}]{Au89}
Au, C.-K., 1989. Effects of transverse photon exchange in helium {Rydberg}
  states: {Corrections} beyond the {Coulomb-Breit} interaction. Phys. Rev. A
  39, 2789--2795.

\bibitem[{Au et~al.(1984)Au, Feinberg, and Sucher}]{AuFeiSuc84}
Au, C.-K., Feinberg, G., Sucher, J., 1984. Retarded long-range interaction in
  {He} {Rydberg} states. Phys. Rev. Lett. 53, 1145--1148.

\bibitem[{Au and Feinberg(1972)}]{AuFei72}
Au, C.-K.~E., Feinberg, G., 1972. Higher-multipole contributions to the
  retarded van der waals potential. Phys. Rev A 6, 2433--2451.

\bibitem[{Babb and Spruch(1987)}]{BabSpr87}
Babb, J.~F., Spruch, L., 1987. Retardation {(Casimir)} effects on high and
  not-so-high {Rydberg} states of helium. Phys. Rev.~A 36, 456--466.

\bibitem[{Babb and Spruch(1994)}]{BabSpr94}
Babb, J.~F., Spruch, L., 1994. Retardation (or {Casimir}) potential for the
  {Rydberg} hydrogen molecule. Phys. Rev. A 50, 3845--3855.

\bibitem[{Barash and Ginzburg(1984)}]{BarGin84}
Barash, Y.~S., Ginzburg, V.~L., 1984. Some problems in the theory of van der
  {W}aals forces. Sov. Phys. Usp. 27, 467--491; Usp. Fiz. Nauk 143 (1984),
  345--389.

\bibitem[{Barton(1999)}]{Bar99}
Barton, G., 1999. Perturbative check on {Casimir} energies of nondispersive
  dielectric spheres. J. Phys. A 32, 525--535.

\bibitem[{Bernab\'eu and Tarrach(1976)}]{BerTar76}
Bernab\'eu, J., Tarrach, R., 1976. Long-range potentials and electromagnetic
  polarizabilities. Ann. Phys. (N.Y.) 102, 323--343.

\bibitem[{Bonin and Kresin(1997)}]{BonKre97}
Bonin, K.~D., Kresin, V.~V., 1997. Electric-dipole polarizabilities of atoms,
  molecules, and clusters. World Scientific, Singapore :River Edge, NJ.

\bibitem[{Bordag et~al.(2009)Bordag, {Klimchitskaya}, Mohideen, and
  {Mostepanenko}}]{BorKliMoh09}
Bordag, M., {Klimchitskaya}, G.~L., Mohideen, U., {Mostepanenko}, V.~M., 2009.
  Advances in the Casimir effect. Vol. 145 of International series of
  monographs on physics. Oxford University Press, Oxford.

\bibitem[{Boyer(1969)}]{Boy69}
Boyer, T.~H., 1969. Recalculations of long-range van der {Waals} potentials.
  Phys. Rev. 180, 19--24.

\bibitem[{Boyer(1974)}]{Boy74}
Boyer, T.~H., 1974. Van der waals forces and zero-point energy for dielectric
  and permeable materials. Phys. Rev. A 9, 2078--2084.

\bibitem[{Buhmann and Welsch(2007)}]{BuhWel07}
Buhmann, S.~Y., Welsch, D.~G., 2007. Dispersion forces in macroscopic quantum
  electrodynamics. Progress in Quant. Electr. 31, 51--130.

\bibitem[{Capasso et~al.(2007)Capasso, Munday, and Iannuzzi}]{CapMunIan07}
Capasso, F., Munday, J.~N., Iannuzzi, D., 2007. Casimir forces and quantum
  electrodynamical {Casimir} forces and quantum electrodynamical torques:
  {Physics} and nanomechanics. IEEE J. Select. Topics Quant. Electron. 13,
  400--414.

\bibitem[{Casimir(1948)}]{Cas48}
Casimir, H. B.~G., 1948. On the attraction between two perfectly conducting
  plates. Proc. K. Ned. Akad. Wet. 60, 793--795.

\bibitem[{Casimir(1987)}]{Cas86}
Casimir, H. B.~G., 1987. Van der {W}aals forces and zero point energy. In:
  Greiner, W. (Ed.), Physics of Strong Fields. Vol. 153 of NATO ASI series.
  Series B, Physics. NATO, Plenum, New York, pp. 957--964.

\bibitem[{Casimir and Polder(1948)}]{CasPol48}
Casimir, H. B.~G., Polder, D., 1948. The influence of retardation on the
  {London-van der Waals} forces. Phys. Rev. 73, 360--372.

\bibitem[{Chen and Chung(1996)}]{CheChu96}
Chen, M.-K., Chung, K.~T., 1996. Retardation long-range potentials between
  helium atoms. Phys. Rev. A 53, 1439--1446.

\bibitem[{{Cirone} and {Passante}(1996)}]{CirPas96}
{Cirone}, M., {Passante}, R., 1996. {Vacuum field correlations and the
  three-body Casimir - Polder potential}. J. Phys. B 29, 1871--1875.

\bibitem[{Compagno et~al.(1995{\natexlab{a}})Compagno, Palma, Passante, and
  {Persico}}]{ComPalPas95}
Compagno, G., Palma, G.~M., Passante, R., {Persico}, F., 1995{\natexlab{a}}.
  Atoms dressed and partially dressed by the zero-point fluctuations of the
  electromagnetic-field. J. Phys. B 28, 1105--1158.

\bibitem[{Compagno et~al.(1995{\natexlab{b}})Compagno, Passante, and
  Persico}]{ComPasPer95}
Compagno, G., Passante, R., Persico, F., 1995{\natexlab{b}}. Atom-field
  interactions and dressed atoms. No.~17 in Cambridge studies in modern optics.
  Cambridge University Press, Cambridge.

\bibitem[{{Compagno} et~al.(2006){Compagno}, {Passante}, and
  {Persico}}]{ComPasPer06}
{Compagno}, G., {Passante}, R., {Persico}, F., 2006. {Edwin {P}ower and the birth
  of dressed atoms}. Contemp. Phys. 47, 269--278.

\bibitem[{Compagno and Salamone(1991)}]{ComSal91}
Compagno, G., Salamone, G.~M., 1991. Structure of the electromagnetic field
  around the free electron in nonrelativistic {QED}. Phys. Rev. A 44~(9),
  5390--5400.

\bibitem[{Craig and Power(1969)}]{CraPow69}
Craig, D.~P., Power, E.~A., 1969. The asymptotic casimir-polder potential from
  second-order perturbation theory and its generalization for anisotropic
  polarizabilities. Int. J. Quant. Chem. 3, 903--911.

\bibitem[{Dalgarno et~al.(1968)Dalgarno, Drake, and Victor}]{DalDraVic68}
Dalgarno, A., Drake, G.~W., Victor, G.~A., 1968. Nonadiabatic long-range
  forces. Phys. Rev. 176, 194--197.

\bibitem[{Deal and Young(1971)}]{DeaYou71}
Deal, W., Young, R., 1971. The long-range retarded interaction between two
  hydrogen atoms. Chemical Physics Letters 11, 385--386.

\bibitem[{Drake(1992)}]{Drake-LRF}
Drake, G. W.~F., 1992. {High-precision calculations for the Rydberg state of
  helium}. In: Levin, F.~S., Micha, D. (Eds.), Long Range Casimir Forces: {T}heory and
  Recent Experiments in Atomic Systems. Plenum Press, New York, pp. 107--217.

\bibitem[{Drake(1993)}]{Dra93}
Drake, G. W.~F., 1993. Energies and asymptotic analysis for helium {R}ydberg
  states. In: Advances in Atomic, Molecular, and Optical Physics. Vol.~31.
  Academic Press, San Diego, pp. 1--62.

\bibitem[{Emig(2008)}]{Emi08}
Emig, T., 2008. Fluctuation-induced quantum interactions between compact
  objects and a plane mirror. J. Stat. Mech. 2008, P04007.

\bibitem[{Emig(2010)}]{Emi10}
Emig, T., 2010. Casimir physics: Geometry, shape and material.
  arxiv.org/abs/1003.0192.

\bibitem[{Emig and Jaffe(2008)}]{EmiJaf08}
Emig, T., Jaffe, R.~L., 2008. Casimir forces between arbitrary compact objects.
  J. Phys. A 41, 164001.

\bibitem[{Feinberg(1974)}]{Fei74}
Feinberg, G., 1974. Retarded dispersion forces between conducting spheres.
  Phys. Rev. B 9, 2490--2496.

\bibitem[{Feinberg and Sucher(1968)}]{FeiSuc68}
Feinberg, G., Sucher, J., 1968. General form of the retarded van der {Waals}
  potential. J. Chem. Phys. 48, 3333--3334.

\bibitem[{Feinberg and Sucher(1970)}]{FeiSuc70}
Feinberg, G., Sucher, J., 1970. General theory of the van der {Waals}
  interaction: {A} model-independent approach. Phys. Rev. A 2, 2395--2415.

\bibitem[{Feinberg and Sucher(1983)}]{FeiSuc83}
Feinberg, G., Sucher, J., 1983. Long-range forces between a charged and neutral
  system. Phys. Rev. A 27, 1958--1967.

\bibitem[{Feinberg et~al.(1989)Feinberg, Sucher, and Au}]{FeiSucAu89}
Feinberg, G., Sucher, J., Au, C.~K., 1989. The dispersion theory of dispersion
  forces. Phys. Rep. 180, 83--157.

\bibitem[{{Henkel} et~al.(2008){Henkel}, {Boedecker}, and
  {Wilkens}}]{HenBoeWil08}
{Henkel}, C., {Boedecker}, G., {Wilkens}, M., 2008. {Local fields in a soft
  matter bubble}. Appl. Phys. B 93, 217--221.

\bibitem[{Hessels(1992)}]{Hes92}
Hessels, E.~A., 1992. Higher-order relativistic corrections to the polarization
  energies of helium {Rydberg} states. Phys. Rev. A 46, 5389--5396.

\bibitem[{Holstein(2008)}]{Hol08}
Holstein, B.~R., 2008. Long range electromagnetic effects involving neutral
  systems and effective field theory. Phys. Rev. D 78, 013001.

\bibitem[{Holstein and Donoghue(2004)}]{HolDon04}
Holstein, B.~R., Donoghue, J.~F., 2004. Classical physics and quantum loops.
  Phys. Rev. Lett. 93, 201602.

\bibitem[{{Jaffe}(2005)}]{Jaf05}
{Jaffe}, R.~L., 2005. {Casimir effect and the quantum vacuum}. Phys. Rev. D
  72, 021301.

\bibitem[{Jenkins et~al.(1994)Jenkins, Salam, and
  Thirunamachandran}]{JenSalThi94b}
Jenkins, J.~K., Salam, A., Thirunamachandran, T., 1994. Retarded dispersion
  interaction energies between chiral molecules. Phys. Rev. A 50, 4767--4777.

\bibitem[{Johnson et~al.(1967)Johnson, Epstein, and Meath}]{JohEpsMea67}
Johnson, R.~E., Epstein, S.~T., Meath, W.~J., 1967. Evaluation of long-range
  retarded interaction energies. J. Chem. Phys. 47, 1271--1274.

\bibitem[{Kelsey and Spruch(1978{\natexlab{a}})}]{KelSpr78b}
Kelsey, E.~J., Spruch, L., 1978{\natexlab{a}}. Retardation effects and the
  vanishing as $r\sim{}\infty{}$ of the nonadiabatic $r^{-6}$ interaction of
  the core and a high-{R}ydberg electron. Phys. Rev. A 18, 1055--1056.

\bibitem[{Kelsey and Spruch(1978{\natexlab{b}})}]{KelSpr78}
Kelsey, E.~J., Spruch, L., 1978{\natexlab{b}}. Retardation effects on high
  {R}ydberg states---retarded $r^{-5}$ polarization potential. Phys. Rev.~A
  18, 15--25.

\bibitem[{Kleinman et~al.(1968)Kleinman, Hahn, and {Spruch}}]{KleHahSpr68}
Kleinman, C.~J., Hahn, Y., {Spruch}, L., 1968. Dominant nonadiabatic
  contribution to the long-range electron-atom interaction. Phys. Rev. 165~(1),
  53--62.

\bibitem[{Levin and Micha(1992)}]{LevMic92}
Levin, F.~S., Micha, D. (Eds.), 1992. Long Range Casimir Forces: Theory and Recent
  Experiments in Atomic System. Plenum Press, New York.

\bibitem[{Lundeen(2005)}]{Lun05}
Lundeen, S.~R., 2005. Fine structure in high-$L$ {Rydberg} states: {A} path to
  properties of positive ions. In: Advances in Atomic, Molecular, and Optical
  Physics. Vol.~52. Elsevier Academic, San Diego, pp. 161--208.

\bibitem[{Luo et~al.(1993)Luo, Kim, Giese, and Gentry}]{LuoKimGie93}
Luo, F., Kim, G., Giese, C.~F., Gentry, W.~R., 1993. Influence of retardation
  on the higher-order multipole dispersion contributions to the helium dimer
  potential. J. Chem. Phys. 99, 10084--10085.

\bibitem[{Mahanty and Ninham(1976)}]{MahNin76}
Mahanty, J., Ninham, B.~W., 1976. Dispersion forces. Academic Press, London.

\bibitem[{Marcus(2009)}]{Mar09}
Marcus, A., October 2009. Research in a vacuum: {DARPA} tries to tap elusive
  {Casimir} effect for breakthrough technology.
  \url{http://www.scientificamerican.com/article.cfm?id=darpa-casimir-effect-research}, web page, published October 12, 2009.

\bibitem[{Marinescu et~al.(1994)Marinescu, Babb, and Dalgarno}]{MarBabDal94}
Marinescu, M., Babb, J.~F., Dalgarno, A., 1994. Long-range potentials,
  including retardation, for the interaction of two alkali-metal atoms. Phys.
  Rev. A 50, 3096--3104.

\bibitem[{Marinescu and You(1999)}]{MarYou99}
Marinescu, M., You, L., 1999. Casimir-polder long-range interaction potentials
  between alkali-metal atoms. Phys. Rev. A 59, 1936--1954.

\bibitem[{Meath and Hirschfelder(1966)}]{MeaHir66}
Meath, W.~J., Hirschfelder, J.~O., 1966. Relativistic intermolecular forces,
  moderately long range. J. Chem. Phys. 44, 3197--3209.

\bibitem[{Milonni(1994)}]{Mil94}
Milonni, P.~W., 1994. The Quantum Vacuum. Academic Press, San Diego.

\bibitem[{Milonni and Lerner(1992)}]{MilLer92}
Milonni, P.~W., Lerner, P.~B., 1992. Extinction theorem, dispersion forces, and
  latent heat. Phys. Rev. A 46, 1185--1193.

\bibitem[{Milton(2001)}]{Mil01}
Milton, K.~A., 2001. The {Casimir} Effect: {Physical} manifestations of
  zero-point energy. World Scientific, Singapore.

\bibitem[{O'Carroll and Sucher(1968)}]{OCaSuc68}
O'Carroll, M., Sucher, J., 1968. Exact computation of the van der {Waals}
  constant for two hydrogen atoms. Phys. Rev. Lett. 21, 1143--1146.

\bibitem[{Pachucki(2005)}]{Pac05}
Pachucki, K., 2005. Relativistic corrections to the long-range interaction
  between closed-shell atoms. Phys. Rev. A 72, 062706.

\bibitem[{Panella and Widom(1994)}]{PanWid94}
Panella, O., Widom, A., 1994. Casimir effects in gravitational interactions.
  Phys. Rev. D 49, 917--922.

\bibitem[{Panella et~al.(1990)Panella, Widom, and Srivastava}]{PanWidSri90}
Panella, O., Widom, A., Srivastava, Y., 1990. Casimir effects for charged
  particles. Phys. Rev. B 42, 9790--9793.

\bibitem[{Parsegian(2006)}]{Par06}
Parsegian, V.~A., 2006. van der {W}aals forces. Cambridge, Cambridge.

\bibitem[{Passante and Power(1987)}]{PasPow87}
Passante, R., Power, E.~A., 1987. Electromagnetic-energy-density distribution
  around a ground-state hydrogen atom and connection with van der waals forces.
  Phys. Rev. A 35, 188--197.

\bibitem[{Piszczatowski et~al.(2009)Piszczatowski, {\L}ach, Przybytek, Komasa,
  Pachucki, and Jeziorski}]{PisLacPrz09}
Piszczatowski, K., {\L}ach, G., Przybytek, M., Komasa, J., Pachucki, K.,
  Jeziorski, B., 2009. Theoretical determination of the dissociation energy of
  molecular hydrogen. J. Chem. Theory Comp. 5, 3039--3048.

\bibitem[{Power and Zienau(1957)}]{PowZie57a}
Power, E., Zienau, S., 1957. On the physical interpretation of the relativistic
  corrections to the van der {W}aals force found by {P}enfield and {Z}atskis.
  Journal of the Franklin Institute 264, 403--407.

\bibitem[{Power and Thirunamachandran(1993)}]{PowThi93b}
Power, E.~A., Thirunamachandran, T., 1993. {Casimir-Polder} potential as an
  interaction between induced dipoles. Phys. Rev. A 48, 4761--4763.

\bibitem[{Power and Thirunamachandran(1996)}]{PowThi96}
Power, E.~A., Thirunamachandran, T., 1996. Dispersion interactions between
  atoms involving electric quadrupole polarizabilities. Phys. Rev. A 53~(3),
  1567--1575.
  
\bibitem[{Przybtek et~al.(2010)Przybytek, Cencek, Komasa, {\L}ach, Jeziorski,
Szalewicz}]{PrzCenKom10}
Przybytek, M., Cencek, W., Komasa, J., {\L}ach, G., Jeziorski, B.,
and 
Szalewicz, K., 2010.
{Relativistic and quantum electrodynamics effects in the helium pair potential}.
Phys. Rev. Lett. 104, 183003.

\bibitem[{Rado\.zycki(1990)}]{Rad90a}
Rado\.zycki, T., 1990. {The electromagnetic virtual cloud of the ground-state
  hydrogen atom---a quantum field theory approach}. J. Phys. A 23, 4911--4923.

\bibitem[{Rugh et~al.(1999)Rugh, Zinkernagel, and Cao}]{RugZinCao99}
Rugh, S.~E., Zinkernagel, H., Cao, T.~Y., 1999. The {Casimir} effect and the
  interpretation of the vacuum. Stud. Hist. Philos. Sci. B: Stud. Hist. Philos.
  Mod. Phys. 30, 111--139.

\bibitem[{{Salam} and {Thirunamachandran}(1996)}]{SalThi96}
Salam, A. and Thirunamachandran, T. 1996. A new generalization
of the Casimir-Polder potential to higher electric multipole polarizabilities.
J. Chem. Phys. 104, 5094--5099.

\bibitem[{Salam(2000)}]{Sal00c}
Salam, A., 2000. Comment on ``{Casimir-Polder} long-range interaction
  potentials between alkali-metal atoms''. Phys. Rev. A 62, 026701.

\bibitem[{Salam(2010)}]{Sal10}
Salam, A., 2010. Molecular quantum electrodynamics. Wiley, Hoboken, NJ.

\bibitem[{{Schaden} and {Spruch}(1998)}]{SchSpr98}
{Schaden}, M., {Spruch}, L., 1998. {Infinity-free semiclassical evaluation
  of Casimir effects}. Phys. Rev. A 58, 935--953.

\bibitem[{{Schwinger}(1975)}]{Sch75}
{Schwinger}, J., 1975. {Casimir effect in source theory}. Lett. Math. Phys. 1,
  43--47.

\bibitem[{Spagnolo et~al.(2007)Spagnolo, Dalvit, and Milonni}]{SpaDalMil07}
Spagnolo, S., Dalvit, D. A.~R., Milonni, P.~W., 2007. van der waals
  interactions in a magnetodielectric medium. Phys. Rev. A 75, 052117.

\bibitem[{Spruch and Kelsey(1978)}]{SprKel78}
Spruch, L., Kelsey, E.~J., 1978. Vacuum fluctuation and retardation effects on
  long-range potentials. Phys. Rev. A 18, 845--852.

\bibitem[{{Spruch} and {Tikochinsky}(1993)}]{TikSpr93a}
{Spruch}, L., {Tikochinsky}, Y., 1993. {Elementary approximate derivations of
  some retarded {Casimir} interactions involving one or two dielectric walls}.
  Phys. Rev. A 48, 4213--4222.

\bibitem[{Stevens and Lundeen(2000)}]{SteLun00}
Stevens, G.~D., Lundeen, S.~R., 2000. Experimental studies of helium {Rydberg}
  fine structure. Comments At. Molec. Phys., Comments Mod. Phys. Part D 1,
  207--219.

\bibitem[{Sucher and Feinberg(1992)}]{SucFei-LRF}
Sucher, J., Feinberg, G., 1992. Long-range electromagnetic forces in quantum
  theory: {Theoretical} formulations. In: Levin, F.~S., Micha, D. (Eds.), Long
  Range Casimir Forces: {T}heory and Recent Experiments in Atomic Systems. Plenum
  Press, New York, pp. 273--348.

\bibitem[{Thirunamachandran(1988)}]{Thi88}
Thirunamachandran, T., 1988. Vacuum fluctuations and intermolecular
  interactions. Phys. Scr. T21, 123--128.

\bibitem[{Voronin et~al.(2005)Voronin, Froelich, and Zygelman}]{VorFroZyg05}
Voronin, A.~Y., Froelich, P., Zygelman, B., 2005. Interaction of ultracold
  antihydrogen with a conducting wall. Phys. Rev. A 72, 062903.

\bibitem[{Watson(1991)}]{Wat91}
Watson, G.~I., 1991. Two-electron perturbation problems and {Pollaczek}
  polynomials. J. Phys. A 24, 4989--4998.

\bibitem[{Wennerstrom et~al.(1999)Wennerstrom, Daicic, and
  Ninham}]{WenDaiNin99}
Wennerstrom, H., Daicic, J., Ninham, B.~W., 1999. Temperature dependence of
  atom-atom interactions. Phys. Rev. A 60, 2581--2584.

\bibitem[{Yan et~al.(1997)Yan, Dalgarno, and Babb}]{YanDalBab97}
Yan, Z.-C., Dalgarno, A., Babb, J.~F., 1997. Long-range interactions of lithium
  atoms. Phys. Rev. A 55, 2882--7.

\bibitem[{Zygelman et~al.(2003)Zygelman, Dalgarno, Jamieson, and
  Stancil}]{ZygDalJam03}
Zygelman, B., Dalgarno, A., Jamieson, M.~J., Stancil, P.~C., 2003. Multichannel
  study of spin-exchange and hyperfine-induced frequency shift and line
  broadening in cold collisions of hydrogen atoms. Phys. Rev. A 67, 042715.

\end{thebibliography}

\end{document}